\documentclass[11pt]{article}
\usepackage{moriond,epsfig,amssymb,amsmath,slashed}

\def\be{\begin{equation}}
\def\ee{\end{equation}}
\def\bea{\begin{eqnarray}}
\def\eea{\end{eqnarray}}

\begin{document}
\vspace*{4cm}
\title{Going Beyond the Minimal Composite Higgs Model}

\author{ Ben GRIPAIOS}

\address{CERN PH-TH, Case C01600, 1211 Geneva 23, Switzerland.}

\maketitle\abstracts{
If electroweak symmetry breaking arises via strong dynamics, electroweak precision tests and flavour physics experiments suggest that the minimal model should closely resemble the Standard Model at the LHC. I describe two directions going beyond the minimal model that result in radically different physics at the LHC.
One direction extends the Higgs sector and the other involves composite leptoquark states.
}
\section{Introduction}
Strong coupling provides a solution of the electroweak hierarchy problem that is natural in the literal sense of the word. That is to say, we already have an example in Nature where a hierarchy, namely the one between the proton mass and, say, the Planck scale, is generated by a strongly-coupled theory, QCD. As such, and given the problems suffered by the only weakly-coupled candidate that can stabilize the hierarchy, namely supersymmetry, strongly-coupled dynamics remains an attractive mechanism for electroweak symmetry breaking.

But strongly-coupled dynamics has severe problems of its own, in the form of clashes with electroweak precision tests and flavour-changing neutral currents.
The former can be solved, to some extent, by clever use of symmetries. To see how this may occur, let me first remark that we do not yet know what the `symmetry' of electroweak symmetry is. Certainly, we do know that it contains the $SU(3) \times SU(2)_L \times U(1)_Y$ of the Standard Model (SM) as a gauged subgroup, but it is quite possible that the true global symmetry of the strongly-coupled sector is somewhat larger. If we enlarge the $SU(2)_L \times U(1)_Y$ to $SU(2)_L \times SU(2)_R \simeq SO(4)$ (which is an accidental symmetry of the renormalizable SM Higgs potential), then we find that the $W$ and $Z$ bosons automatically obtain their measured mass ratio.\cite{Sikivie:1980hm} Furthermore, by adding the discrete parity that interchanges $L \leftrightarrow R$ (or equivalently enlarging $SO(4)$ to $O(4)$), we can suppress 
unwanted corrections to the coupling $Z b \overline{b}$.\cite{Agashe:2006at} The remaining nuisance is the $S$-parameter, which, alas, no symmetry can forbid without simultaneously forbidding electroweak symmetry breaking. Nevertheless, one can still use symmetry to argue that $S$, although it cannot vanish, could be small.

The argument goes as follows. If $SU(2)_L$ were a symmetry of the vacuum, then $S$ would indeed vanish.\cite{Inami:1992rb} This implies that $S$ must be proportional to some positive power of the electroweak vev, $v$; in fact, since $S$ transforms as an $SU(2)_L$ triplet, whilst $v$ transforms as a doublet, we have that $S \propto v^2$. Now, $S$ is dimensionless, so we must have that $S \simeq v^2/f^2$, where the scale $f$ is set by the strong dynamics; if we could arrange for $v/f$ to be somewhat less than unity, by some dynamical accident, then we might end up with an acceptably-small value for $S$. One way to do this is to further enlarge the global symmetry of the strong sector to $SO(5)$ and then to decree that strong dynamics breaks it to $SO(4)$.\cite{Agashe:2004rs} The theory then contains four pseudo-Nambu-Goldstone bosons (PNGBs), which have the quantum numbers of the SM Higgs. $v$ is then set by minimizing the potential of these PNGBs, which is generated by loop effects involving gauge and (proto-)Yukawa couplings (which break the global symmetry) and wherein the hoped-for miracle might occur. 

As for the problems with flavour-changing neutral currents (FCNCs), these too can be ameliorated, by thinking carefully about the way in which fermion masses arise in a strongly-coupled theory. The obvious way to generate fermion masses is to declare that, as in the SM, fermion masses arise via elementary fermions (external to the strong sector) being coupled bi-linearly to a scalar operator of the strong sector that carries the same SM quantum numbers as the SM Higgs. Schematically,
\begin{gather} \label{old}
\mathcal{L} \supset   \frac{\overline{q} \mathcal{O}_{H} u }{\Lambda^{d-1}},
\end{gather}
where I have allowed for the operator $\mathcal{O}_H$ to have arbitrary dimension $d>1$. Now, this operator certainly allows for the observed fermion masses to be generated. However, to this effective Lagrangian we should also add other operators that are compatible with the symmetries of the theory. Amongst these are
\begin{gather} \label{danger}
\mathcal{L} \supset \frac{\overline{q} q \overline{q} q}{\Lambda^{2}} + \Lambda^{4 - d'}\mathcal{O}_{H}^\dagger \mathcal{O}_{H}.
\end{gather}
The first of these is responsible for flavour changing neutral currents; for these to be small enough, $\Lambda > 10^{3-4}$ TeV. But then, in order to get a mass as large as that of the top from the operator in eq. \ref{old} (which is suppressed by $\Lambda$), we need to choose $d$ to be rather small: $d \lesssim 1.2 - 1.3$.\cite{Luty:2004ye} Next, we need to worry about the second operator in eq. \ref{danger}. In order not to de-stabilize the hierarchy, its dimension, $d'$, had better be greater than four, rendering it irrelevant.\footnote{It is, perhaps, instructive to see how the hierarchy problem of the SM is cast in this language. There, $\mathcal{O}_{H}$ corresponds to the Higgs, with dimension close to unity, whilst $\mathcal{O}_{H}^\dagger \mathcal{O}_{H}$ is the Higgs mass operator, with dimension close to two.} So what is the problem? The limit in which $d \rightarrow 1$ corresponds to a free theory, and in that limit $d' \rightarrow 2d$. So in order to have an acceptable theory, we need a theory containing a scalar operator $\mathcal{O}_H$ (with the right charges) with a dimension that is close to the free limit, but such that the theory is nevertheless genuinely strongly-coupled, with the dimension of $\mathcal{O}^\dagger_H \mathcal{O}_H$ greater than four.

In fact, this possibility is not so outlandish as may first appear. Indeed, the Ising model in two dimensions has operators with similar properties.\cite{Rattazzi:2008pe} But sadly, in the real world of four dimensions, we are close to a theorem stating that such a theory cannot exist.\cite{Rattazzi:2008pe,Rychkov:2009ij} 

One way out of this impasse is to note that there exists another way to generate fermion masses.\cite{Kaplan:1991dc} In this alternative, fermion masses arise by allowing the elementary fermions to couple linearly to fermionic operators of the strong sector. Schematically, the Lagrangian is
\begin{gather} \label{mix}
\mathcal{L} \sim   \overline{q} \mathcal{O}_{q} +    \overline{u} \mathcal{O}_{u} +   \overline{\mathcal{O}}_{q} \mathcal{O}_{q} + \overline{\mathcal{O}}_{u} \mathcal{O}_{u} +  \overline{\mathcal{O}}_{q} \mathcal{O}_H \mathcal{O}_{u}
\end{gather}
and the light fermion masses arise by mixing with heavy fermionic resonances of the strong sector, which feel the electroweak symmetry breaking. The beauty of this mechanism is that fermion masses can now be generated by relevant operators ({\em cf.} the operator that generates masses in eq. \ref{old}, which is at best marginal, since $d>1$); this means that one can, in principle, send $\Lambda$ to infinity and the problem of flavour changing neutral currents can be completely decoupled. There is even a further bonus, in that the light fermions of the first and second generations, which are the ones that flavour physics experiments have most stringently probed, are the ones that are least mixed with the strong sector and the flavour-changing physics that lies therein.

In conclusion, the problems of electroweak precision tests and flavour-changing neutral currents may be avoided in a strongly-coupled model. But lest the reader be tempted to leap out of his or her bathtub, declaring electroweak symmetry breaking a closed subject, it is perhaps pertinent to remark that we do not actually have an explicit model in hand. We have shown only that such a model, if one exists, is not necessarily inconsistent.

Nevertheless, it is interesting to peruse what the implications for the LHC might be. We have argued that the minimal model should have $SO(5)$ symmetry broken to $SO(4)$, with the SM subgroup gauged. The light degrees of freedom of such a model are just those of the SM, and the constraints on the $S$ parameter (and also those from FCNCS) suggest that $v/f$ should be rather small. This implies not only that the strongly-coupled resonances must be rather heavy (multi-TeV), but also that the couplings of the light degrees of freedom should be close to their SM values (the departures being set by $v/f$). As such, the minimal model will look very much like the SM at the LHC.

To find a strongly-coupled model which is compliant with electroweak precision and flavour tests, and whose predictions differ from those of the SM as regards the LHC, we need to go beyond the minimal model. In the following, I describe two possible directions.

\section{Beyond the Minimal Composite Higgs Model}
One possible direction is to enlarge the symmetry of the strong sector, whilst retaining the custodial symmetries and SM Higgs degrees of freedom described above.\cite{Gripaios:2009pe}
An obvious choice is to enlarge the global symmetry from $SO(5)$ to $SO(6)$. As for the unbroken subgroup, one could stick with the original breaking to $SO(4)$, or enlarge it to $SO(4)\times SO(2)$, or $SO(5)$. Of these three possibilities, the first two give rise to models with two Higgs doublets, and these re-introduce a problem with the $T$-parameter. To wit, the custodial symmetry $SO(4)$ only protects $T$ if an $SO(3)$ is preserved in the vacuum. With a single Higgs doublet this is guaranteed, but with two Higgs doublets the symmetry will, in general, be further broken to $SO(2)$.\cite{Katz:2005au} The third possibility, by contrast, contains only an extra SM singlet. This extra singlet does not spoil the agreement with existing experiments, but it can dramatically change the phenomenology at the LHC. 

Most strikingly, the singlet, being a PNGB, can naturally be rather light. Indeed, the singlet does not get a contribution to its mass from gauge interactions and the dynamics of the strong sector may be such that the contribution to the mass of the singlet coming from couplings to fermions may also be rather small. In particular, there is a $U(1)$ symmetry of the strong sector that can be used to make the singlet arbitrarily light in a way that is technically natural.\footnote{This is in contrast with some other models that advocate non-standard decays of the Higgs, in which symmetries purporting to guarantee the lightness of the singlet are at best approximate.} As such, the singlet can have a mass in the $1-10$ GeV range, in which case the Higgs decays dominantly into a pair of singlets, which each in turn decay into 
two SM particles. Which SM particles they decay to depends on the mass and couplings of the singlet: the parameter space of the model in fact allows many possibilities, including $b\overline{b}$, $\tau \overline{\tau}$, $c \overline{c}$, such that a Higgs with mass below 114 GeV could have been missed at LEP. For the latest bounds, see ref. \cite{Schael:2010aw}.

There are a number of other interesting phenomenological possibilities. One is that, if the singlet obtains a vev, one has an additional source of spontaneous $CP$ violation in the theory. This may be of relevance for electroweak baryogenesis. Another is that, in this model, there may be physics associated with the Wess-Zumino-Witten term \cite{Wess:1971yu,Witten:1983tw} that captures topological information about the strongly-coupled sector.\cite{Gripaios:2007tk,Gripaios:2008ei}
\section{Composite Leptoquarks}
To see another way in which strongly-coupled models might deviate from the SM,\cite{Gripaios:2009dq} let us look again at the generation of fermion masses.
If fermion masses are generated according to eq. \ref{mix} (as suggested by FCNCs), then the strong sector must know about $SU(3)$ colour, in that the strong sector must
possess fermionic operators, such as those that couple to elementary quarks, that carry colour (as well as electroweak charges)
This is fundamentally different to the fermion mass generation mechanism of eq.  \ref{old}, where the strong sector need only know about the $SU(2)\times U(1)$ part of the SM gauge symmetry.

Now, given that the theory necessarily contains fermionic operators (or rather, light resonances) carrying colour and electroweak charges, it is plausible to imagine that the strong
sector might also feature light, bosonic resonances carrying colour and electroweak charges. Such operators could couple to a quark and a lepton, playing the r\^{o}le of leptoquark states. 

As an explicit example, let us take a toy model where the strongly-coupled dynamics is just an $SU(3)$ gauge theory, like QCD. Indeed, the original model \cite{Kaplan:1991dc} was of just this form and there one finds, as well as baryonic resonances that are necessary for fermion mass generation as per eq. \ref{mix}, mesonic resonances that are leptoquarks.
As another example, we could also consider composite models which incorporate gauge coupling unification by enlarging the global symmetries of the strong sector to some GUT group (see ref. \cite{Agashe:2005vg} and refs. therein). There, bosonic resonances, like the Higgs, must come in complete representations of the GUT group, and so leptoquark states seem inevitable.  

What properties might we expect these composite leptoquarks to have? Generically, we expect resonances around the TeV scale, and these may or may not be within the reach of the LHC. But leptoquarks may also be much lighter, if they arise (like the Higgs) as PNGBs of the strong dynamics. Then, na\"{\i}ve dimensional analysis suggests that a mass of around several hundred GeV seems more likely.\cite{Gripaios:2009dq}

In other leptoquark scenarios (such as plain-Vanilla GUTs), such light leptoquarks would have been ruled out my many orders of magnitude by bounds from various flavour-changing physics processes to which they contribute. For a comprehensive list, see ref. \cite{Davidson:1993qk}. But in our scenario, leptoquark contributions to flavour-changing processes
are suppressed by exactly the same mechanism that suppresses generic strong-sector contributions to flavour physics: the light generations, which are most stringently probed in experiments, are the ones least mixed with the strong sector in general, and the leptoquarks in particular.

Roughly speaking, the leptoquark couplings scale hierarchically like the Yukawa couplings, since they arise from the same mixings. Using a plausible estimate for the leptoquark couplings, it was found in \cite{Gripaios:2009dq} that all the bounds in ref. \cite{Davidson:1993qk} were evaded by at least a factor of ten, except for the bounds coming from the decays $\mu \rightarrow e \gamma$ and $\tau \rightarrow \mu \gamma$. Here, the constraints can only be satisfied if the leptoquarks are chiral: that is, they couple only to either left- or right-handed quarks, but not to both. A glance at the classification table of leptoquarks in the PDG shows that this rules out only a small fraction of the possible leptoquarks. In the case that leptoquarks are chiral, the bounds from the aforementioned processes are also a factor of ten or more above my estimates.

What about the prospects for the LHC? The composite leptoquarks described here only have dominant couplings to third generation quarks and leptons. As a result, they will be only pair-produced at the LHC, via QCD interactions. These pair-produced leptoquarks will then each decay to a $b$ or a $t$ and a $\tau$ or a $\nu_\tau$,
so the dominant final states will be $2b2\tau$, $2b +\slashed{E}_T$, $2t2\tau$, or $2t +\slashed{E}_T$. The leptoquarks are now being added to the HERWIG event generator, in preparation for future phenomenological studies.\cite{future}

\section{Summary}
If strong dynamics is responsible for electroweak symmetry breaking, it seems that the minimal incarnation thereof will look a lot like the SM at the LHC.
I presented two directions in which one can obtain non-minimal models, which differ spectacularly from the SM in their LHC predictions. One direction is to extend the Higgs sector, and the other involves composite, third-generation leptoquark states at the TeV scale.
\section*{Acknowledgments}
I thank the organizers for their kind invitation to come to Moriond and for an enjoyable conference.

\section*{References}

\end{document}